\documentclass[10pt,journal,compsoc]{./IEEEtran}
\usepackage{booktabs} 
\usepackage[ruled,linesnumbered,vlined]{algorithm2e}
\usepackage[ruled]{algorithm2e} 
\usepackage{multirow}

\usepackage{graphicx}
\usepackage[T1]{fontenc} 
\usepackage{amsmath} 
\usepackage[cmintegrals]{newtxmath} 
\usepackage{bm} 
\usepackage{cite}
\usepackage[switch]{lineno}
\usepackage[marginal]{footmisc}
\usepackage{color}

\usepackage{enumitem}

\usepackage{ifpdf}
\usepackage{ifpdf}
\usepackage{algorithmic}

\usepackage{array}

\ifCLASSOPTIONcompsoc
 \usepackage[caption=false,font=footnotesize,labelfont=sf,textfont=sf]{subfig}
\else
 \usepackage[caption=false,font=footnotesize]{subfig}
\fi
\usepackage{fixltx2e}
\usepackage{dblfloatfix}

\usepackage{url}
\begin{document}
\title{Personalized Graph Neural Networks with Attention Mechanism for Session-Aware Recommendation}
\author{Mengqi~Zhang, Shu~Wu,  Meng Gao,
         Xin Jiang, Ke Xu, and~Liang~Wang,\IEEEmembership{Fellow,IEEE}
}

\markboth{IEEE TRANSACTIONS ON KNOWLEDGE AND DATA ENGINEERING,VOL. 31,NO. 9,bSEPTEMBER 2019}%
{Shell \MakeLowercase{\textit{et al.}}: Bare Demo of IEEEtran.cls for Computer Society Journals}
%


\IEEEtitleabstractindextext{%
\begin{abstract}
The problem of session-aware recommendation aims to predict users' next click based on their current session and historical sessions. Existing session-aware recommendation methods have defects in capturing complex item transition relationships. Other than that, most of them fail to explicitly distinguish the effects of different historical sessions on the current session. To this end, we propose a novel method, named Personalized Graph Neural Networks with Attention Mechanism (A-PGNN) for brevity. A-PGNN mainly consists of two components: one is Personalized Graph Neural Network (PGNN), which is used to extract the personalized structural information in each user behavior graph, compared with the traditional Graph Neural Network (GNN) model, which considers the role of the user when the node embeddding is updated. The other is Dot-Product Attention mechanism, which draws on the Transformer net to explicitly model the effect of historical sessions on the current session. Extensive experiments conducted on two real-world data sets show that A-PGNN evidently outperforms the state-of-the-art personalized session-aware recommendation methods.
\end{abstract}

\begin{IEEEkeywords}
Graph Neural Networks, Attention, Session-Aware Recommendation
\end{IEEEkeywords}}

\maketitle

\IEEEdisplaynontitleabstractindextext

%
\IEEEpeerreviewmaketitle

\section{Introduction}
\IEEEPARstart
With the rapid growth of the amount of information on the Internet, recommender systems have become a fundamental technique to help users alleviate the problem of information overload. Usually, the user's interactions with the system within a given time frame are organized into a session. Predicting the next interaction for anonymous sessions is called session-based recommendation. Generally, the users' identifications and past behaviors can be utilized for the next-click prediction, which is called session-aware recommendation. In this scenario, users' recent behaviors in the current session often reflect their short-term preference, whereas historical session sequences imply the evolution of their long-term preferences over time. Combining the short- and long-term preferences of users have become a vital issue in the session-aware recommendation.

In recent years, most session-aware recommendation studies are conducted on the methods of session-based recommendation. Numerous researches based on deep learning \cite{DBLP:journals/corr/HidasiKBT15,tan2016improved,hidasi2018recurrent} applied Recurrent Neural Networks (RNNs) in session-based recommendation scenarios and have obtained promising results. Attention networks are also a powerful tool to capture user interest in each session\cite{Liu:2018:SSA:3219819.3219950, ying2018sequential}. 
Recently, graph-based models have gained increasing attention. SR-GNN \cite{wu2019session} is the first to apply graph neural networks to capture complex item transition relationships in each session. However, these abovementioned session-based models can only leverage the current anonymous session to make the recommendation. Therefore, the session-aware recommenders came into being\cite{twardowski2016modelling,wu2017session,anelli2017analysis}. To capture the user's interest drift across sessions, recent session-aware works \cite{quadrana2017personalizing,Ruocco2017Inter} use a hierarchical RNN to capture the flow of user interest within and across sessions. The recently proposed HierTCN \cite{you2019hierarchical} employs a hierarchical architecture that contains GRU and Temporal Convolutional Network to capture both the long-term interests and short-term interactions.
\begin{figure}[t]
	\centering
	\subfloat[User interaction sessions]{\includegraphics[scale=0.45]{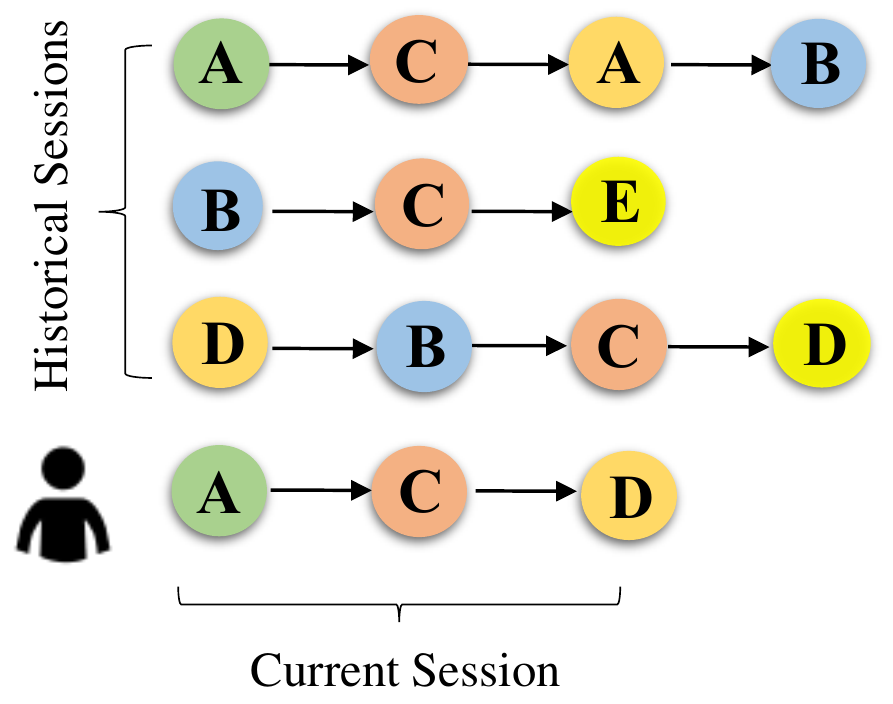}
		\label{fig_first_case}}
	\subfloat[User behavior graph]{\includegraphics[scale=0.45]{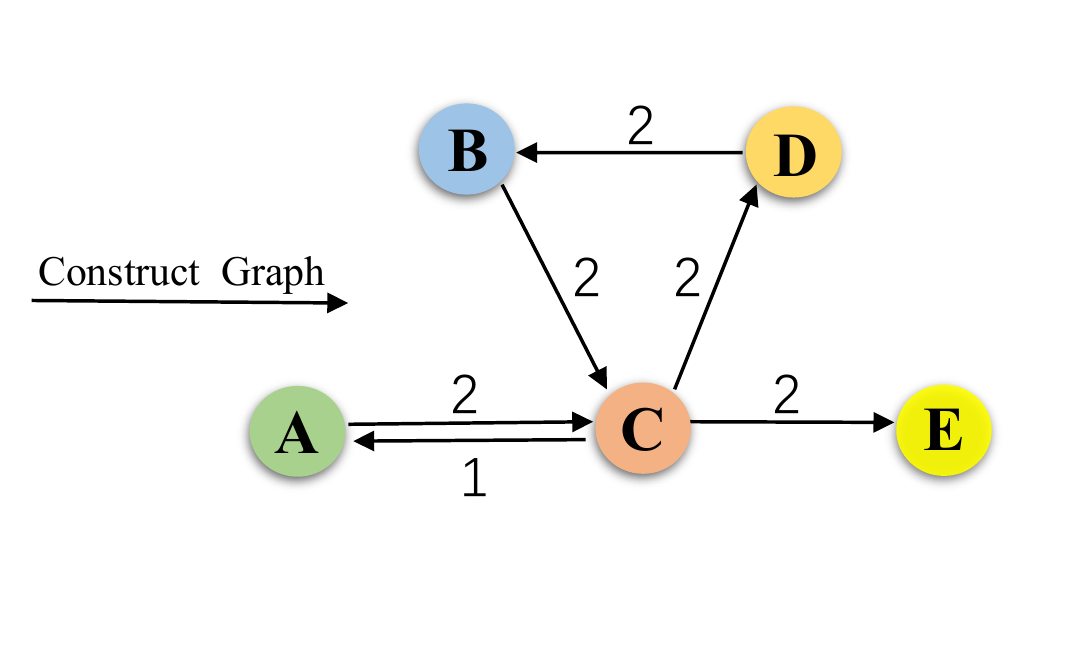}
		\label{fig_second_case}}
	\caption{Construction of user behavior graph. In (a), user's interaction sessions includes historical sessions and current session. (b) is a schematic diagram of the user behavior graph. The number on the directed edge represents the number of times it appears.}
	\label{fig2}
\end{figure}

Despite their effectiveness, we argue that these personalized models remain as the two critical limitations. First, personalized models have defects in capturing complex item transition relationships. Take Figure \ref{fig2} as an example. Figure \ref{fig2} (a) represents all the interaction sessions of user $u$. Based on the transitions between items, the whole session can be converted into a graph as Figure \ref {fig2}(b), where $B$, $C$, and $D$ compose a strongly connected component, which reflects their dense link relationships. Most existing sequence-based session-aware methods are challenging to capture that intricate pattern within and across sessions.
Second, some session-aware methods fail to explicitly distinguish the effects of different historical user sessions on the current session. As a motivating example, suppose a user previously browsed or clicked a digital camera on a shopping site, and his current clicked items are SD and Micro-SD cards. In this case, there is a strong relationship between the historical session and the current session item. If his current interaction is with automotive products, the previously clicked items and the automotive fall into two unrelated categories, which shows that the historical sessions have a minor effect on the current session. However, recent work HierTCN and H-RNN encode historical user sessions into the initial representation of session-level TCN or RNN to assist in making predictions. They all ignore the fine-grained impact of historical sessions on the current session, making the model insufficient to utilize historical information.
To address these limitations, we consider improving the construction of session-graph in SR-GNN model \cite{wu2019session}: building a personalized graph according to both user's current and historical sessions, as shown in Figure \ref {fig2}(b), called user behavior graph. To further reinforce associations between different sessions of each user, we design a personalized graph neural network (PGNN) that considers the role of the user when the node embedding is updated. Furthermore, we use the attention mechanism of Transformer \cite{vaswani2017attention} to model the explicit effect of historical session on each item of the current session. 

To sum up, in this article, we propose a novel method Personalized Graph Neural Networks with Attention Mechanism (A-PGNN). It contains two main components: PGNN and Dot-Product Attention mechanism. We first convert all sessions of each user into a graph, and then feed it into PGNN and Dot-Product net in sequence. Figure \ref{fig:workflow} illustrates the workflow of the proposed A-PGNN model. The details are introduced in Section \ref{proposed model}. Extensive experiments conducted on real-world representative data sets demonstrate the effectiveness of the proposed method over the state-of-the-art methods. The main contributions of this work are summarized as follows:

\begin{itemize}[leftmargin=*]
	\item We design a new graph neural network PGNN for personalized recommendation scenario, which is able to capture complex item transitions in user-specific fashion.

	\item We use the attention mechanism to explicitly model the effect of the user's historical interests on the current session, which shows the superiority of our model in the session-aware recommendation task.

	\item We conduct empirical studies on two real-world data sets. Extensive experiments demonstrate the effectiveness of our proposed model and the contribution of each component.
\end{itemize}

{\section{Related Work}
	
	\subsection{Session-based recommendation}
	Matrix factorization \cite{mnih2007probabilistic, koren2009matrix, koren2011advances} is a general approach used in recommendation systems. The basic objective is to factorize a user-item rating matrix into two low-rank matrices, and each of them represents the latent factors of users or items. The item-based neighborhood methods \cite{Sarwar2001} are a natural solution, in which item similarities are calculated on the co-occurrence in the same session. These methods have difficulty in considering the sequential order of items and generate prediction merely based on the last click. Then, the sequential methods based on Markov chains are proposed, which predict the users' next behavior based on the previous ones. Treating recommendation generation as a sequential optimization problem, \cite{Shani:2002:MRS:2073876.2073930} uses Markov decision processes (MDPs) for the solution. Via factorization of the personalized probability transition matrices of users, FPMC \cite{rendle2010factorizing} models sequential behavior between every two adjacent clicks and provides a more accurate prediction for each sequence. The main drawback of Markov-chain-based models is that they combine past components independently. Such an independence assumption is too strong, and thus confines the prediction accuracy.
	
	Recently, deep neural networks have become the most successful methods in modeling sequence,  such as machine translation\cite{mikolov2013distributed, mikolov2010recurrent, cho2014learning}, conversation machine\cite{serban2016building}. For session-based and sequential recommendation,  the work of \cite{DBLP:journals/corr/HidasiKBT15} proposes the recurrent neural network approach, and then extends to an architecture with parallel RNNs \cite{Hidasi:2016:PRN:2959100.2959167}, which could model sessions based on the clicks and features of the clicked items. After that, some work are proposed based on these RNN methods. An improved RNN \cite{tan2016improved} enhances the performance of recurrent model by using proper data augmentation techniques and taking temporal shifts in user behavior into account. The work of \cite{Jannach:2017:RNN:3109859.3109872} combines the recurrent method with the neighborhood-based method together to mix the sequential patterns as well as the co-occurrence signals. What is more, convolutional neural networks are also used in sequential recommendations to incorporate session clicks with content features\cite{Tuan:2017:CNS:3109859.3109900}. 
	
	Furthermore, a neural attentive recommendation machine with an encoder-decoder architecture, that is, NARM \cite{Li:2017:NAS:3132847.3132926}, utilizes the attention mechanism on RNN to capture the users' features of sequential behavior and main purposes.  SHAN model \cite{ying2018sequential} uses a two-layer hierarchical attention network, which takes the long- and short-term preferences into account. Then, a short-term attention priority model (STAMP) \cite{Liu:2018:SSA:3219819.3219950} using a novel attention memory network, is proposed to efficiently capture both the users' general interests and current interests. 
	However, these abovementioned session-based or sequential models can only leverage the current anonymous session or single sequence to make the recommendation.
	
	\subsection{Session-aware recommendation}
	In session-aware recommendation scenarios, the user behavior in past sessions might provide valuable information for providing recommendations in the next session. In \cite{twardowski2016modelling},  RNN-based approaches are proposed, which leverage additional item features to enhance recommendation capacity. A list-wise deep neural network \cite{Wu:2017:SIE:3132847.3133163} models the limited user behavior within each session and uses a list-wise ranking model to generate the recommendation for each session. The work of \cite{phuong2019neural} proposes some strategies to integrate user expression with RNN models. However, they all fail to effectively use the user's historical information. To this end, the work\cite{quadrana2017personalizing} uses a hierarchical RNN to capture users' short- and long-term preferences for personalized session-based recommendation. Analogous to model Hierarchical RNN, II-RNN model \cite{Ruocco2017Inter} also utilizes multiple RNN to model interest relationships within current and historical sessions.  DANN \cite{Tianan2019Personalizing} exploits a dual attentive neural network to model user's personalized preference and primary purpose in his all sessions. 
	The recently proposed HierTCN \cite{you2019hierarchical} employs a hierarchical architecture that contains GRU and Temporal Convolutional Network to capture both the long-term interests and short-term interactions. However, their encoding mechanism  limits the model's capabilities. In addition, it is difficult to fully capture the complex patterns of user behavior by relying solely on the sequence relationship of the session.

	\subsection{Graph neural networks}
	Nowadays, neural network has been used for generating representation for graph-structured data, for example, social network and knowledge bases. On one hand, extending the word2vec \cite{mikolov2013distributed}, an unsupervised algorithm DeepWalk \cite{Perozzi:2014:DOL:2623330.2623732} is designed to learn representations of graph nodes based on random walk. Following DeepWalk, unsupervised network embedding algorithms LINE \cite{Tang:2015:LLI:2736277.2741093} and node2vec \cite{Grover:2016:NSF:2939672.2939754} are most representative methods. On the other hand, the classical neural network CNN and RNN are also deployed on graph-structured data. Duvenaud et al. \cite{Duvenaud:2015:CNG:2969442.2969488} introduce a convolution neural network that operates directly on graphs of arbitrary sizes and shapes. A scalable approach \cite{DBLP:journals/corr/KipfW16} chooses the convolution architecture via a localized approximation of spectral graph convolutions, which is an efficient variant, and it could operate on graphs directly as well. However, these methods can only be implemented on undirected graphs. Previously, in the form of recurrent neural networks, Graph Neural Networks (GNNs) \cite{1555942} \cite{4700287} are proposed to operate on directed graphs. As a modification of GNN, Gated Graph Neural Networks \cite{DBLP:journals/corr/LiTBZ15} uses gated recurrent units and employs back-propagation through time (BPTT) to compute gradients. Graph Attention Networks (GAT) \cite{DBLP:conf/iclr/VelickovicCCRLB18} applies the attention mechanism to learn the weight of nodes and neighbor nodes. Recently GNN is broadly applied for the different tasks, for example, script event prediction \cite{EEG2018}, situation recognition \cite{8237710}, recommender system\cite{wu2019session} and image classification \cite{8099493}. 
	
	GNN has advantages in processing graph structure data and can be used to capture more abundant information in sequence data. SR-GNN \cite{wu2019session} is the first model to utilize the Gated Graph Neural Networks to capture the complex item transition relationships in session-based recommendation scenarios, but it ignores the role of user in item transition relationship, and fails to use user historical session information to improve recommendation performance. In this work, we propose a model A-PGNN based on improved GGNN, which is more suitable for personalized session-based recommendation scenarios.

\section{The Proposed Method}
\label{proposed model}
In this section, we introduce the proposed A-PGNN\footnote{https://github.com/CRIPAC-DIG/A-PGNN} which applies personalized graph neural networks along with attention mechanism for session-aware recommendation. First, the problem is formulated. Then, we introduce the overview of our proposed method and give more details.
\begin{figure*}[h]
	\centering
	\includegraphics[scale=0.6]{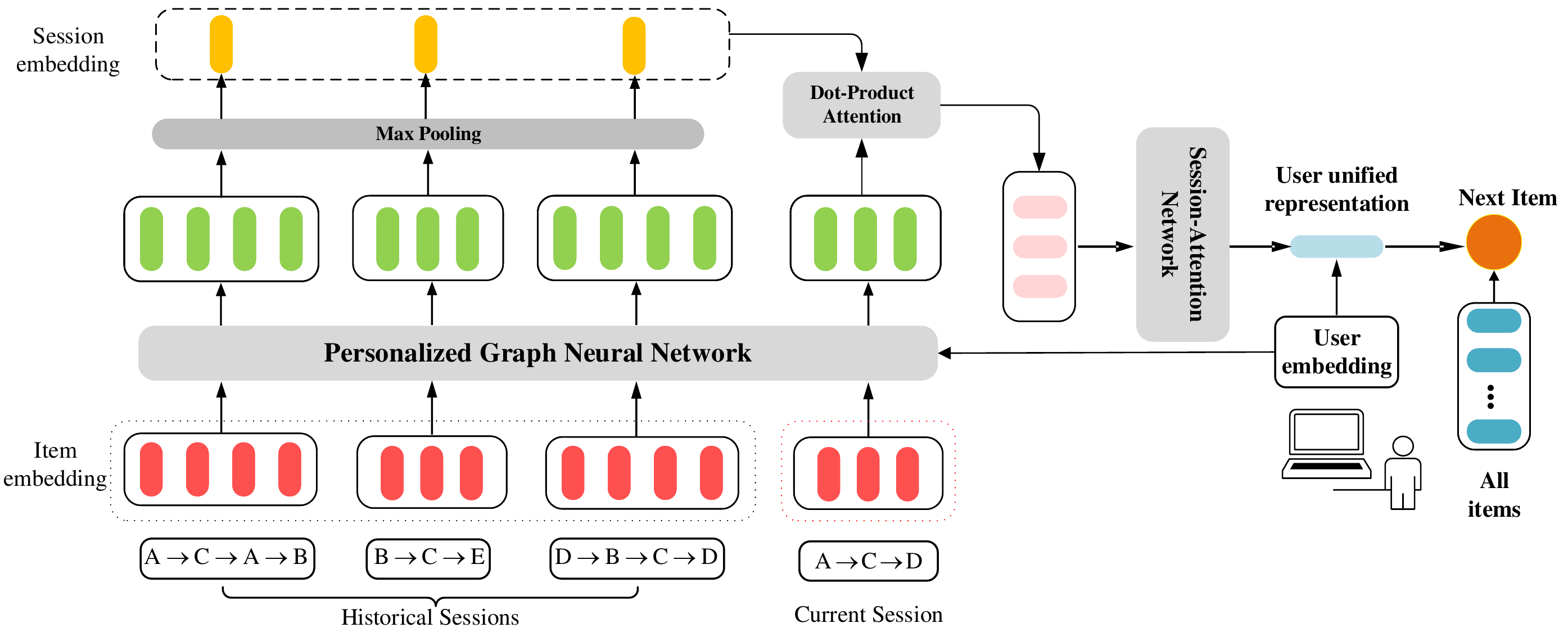}
	\caption{The framework of A-PGNN. Based on the user's all sessions, we first construct a user behavior graph. We then input the user behavior graph along with item and user embedding into PGNN to obtain the item representations. Next, we utilize the max-pooling layer to get the historical session embedding. Following this, the representation of the current session that incorporates the user's historical preferences can be obtained by the Dot-product attention component. Finally, we utilize the Session-Attention network to generate the user's dynamic representation and combine it with the user's static embedding to get a unified user representation for final prediction.}
	\label{fig:workflow}
\end{figure*}
\subsection{Problem Formulation}
\label{sec:problem-formulation}
Let $V = \{v_i\}_{i=1}^{|V|}$ and $U = \{u_i\}_{i=1}^{|U|}$ be the set of items and users in the system, respectively. We describe item $v_i$ with embedding vector $\mathbf{e}_{v_i}\in{\mathbb{R}^d}$ and user $u_i$ with embedding vector $\mathbf{e}_{u_i} \in \mathbb{R}^{d'}$. $d$ and $d'$ are the dimensions of item and user embedding, respecively. For each user $u$, each session $S_i^u={\{v_{i,j}\}_{j=1}^{m_i}} \in \mathcal{S}^u$ represents the sequence in the order of time occurrence, $v_{i,j} \in V$ represents an interactive item of the user within the session $S_i^u$, and $m_i$ is the number of items in session ${S}_i^u$. All sessions of user $u$ can be represented as $\mathcal{S}^u=\{S_i^u\}_{i=1}^{n_u}$, where $n_u$ stands for the total number of sessions for a user $u$,  for convenience, $n_u$ is abbreviated as $n$ and  $\mathcal{S}^u=\{S_i^u\}_{i=1}^{n}$ . Sessions and items are all ordered by timestamps. In general, the last interacted session, that is, $\{S_n^u\}$, is called current session. Other sessions that belong to $\{S_i^u\}_{i=1}^{n-1}$ are called historical session.
Current session and historical session are denoted as $\mathcal{S}_\text{c}^u, \mathcal{S}^u_\text{h}$ respectively.  Given users' all sessions $\mathcal{S}^u$, the goal of session-aware recommendation is to predict the next interactive item $v_{n, {m+1}}$ of the current session $\mathcal{S}_\text{c}^u$. The notations used throughout this paper are summarized in Table \ref{tab:notations}.
\begin{table}	\centering
	\caption{Important notations}
	\resizebox{0.9\columnwidth}{!}{
		\begin{tabular}{cl}
			\toprule
			{\bf Notation} & {\bf Description} \\
			\hline \hline
			$V$							& the set of items \\
			$U$							& the set of users \\
			\hline \hline
			$\mathcal{S}^u$             & user $u$'s all sessions set \\
			$S_i^u$						& the $i$-th session of user $u$'s all sessions \\
			$\mathcal{S}_\text{h}^u$    & the historical sessions of user $u$ \\
			$\mathcal{S}_\text{c}^u$				& the current session of user $u$ \\
			\hline \hline
			$\mathcal{G}^u$			    & the user behavior graph of user $u$\\
			$\mathbf{e}_{v_i}$					& the embedding of item $v_i$ \\
			$\mathbf{e}_{u_i}$						& the embedding of user $u_i$ \\
			$\mathbf{z}_u$			    & the unified representation of user $u$ \\
			\bottomrule
		\end{tabular}
	}
	\label{tab:notations}\end{table}
\subsection{Overview}
Figure \ref{fig:workflow} is the overview of our proposed method. For each user $u$, all sessions $\mathcal{S}^u$ can be modeled as a user behavior graph $\mathcal{G}^u$ (Section \ref{sec:session-graph-modeling}). Then, $\mathcal{G}^u$ is fed into Personalized Graph Neural Network (PGNN) (Section \ref{sec:generate-item-user-embedding}) to capture transitions of items with respect to user $u$. After that, we use the max-pooling layer to get the session embedding, therefore, the historical session embedding matrix can be obtained. Then, we evaluate the explicit impact of the historical session on the current session through Dot-Product attention layer (Section \ref{subsec:attention}). Thereby, we can get user's dynamic interest representations and concatenate it with users' embedding to obtain a unified representation (Section \ref{subsec:user_rep}). Using the representations, we output probability $\hat{\mathbf{y}}$ for all candidate items, where the element $y_i \in \hat{\mathbf{y}}$ is the recommendation score of the corresponding item $v_i \in V$ (Section \ref{sec:make-recommendation}). The items with top-$k$ values will be the candidate items for recommendation. 
\subsection{User Behavior Graph}
\label{sec:session-graph-modeling}
To fully capture the complex item transitions of each user, inspired by SR-GNN \cite{wu2019session}, we construct graph $\mathcal{G}^u$ for each user. As shown in Fig \ref{fig2} (a) and (b), for each user $u$, we model all of his/her sessions $\mathcal{S}^u$ as a directed graph $\mathcal{G}^u = \left(\mathcal{V}^u, \mathcal{E}^u\right)$. In each user behavior graph $\mathcal{G}^u$, node $i$ represents an item $v_{i} \in V$ that user $u$ interacted with. The edge $v_j \to v_i$ represents a user interacts item $v_i$ after $v_j$ in one of his sessions. For this case, we assume that the effect of $v_i$ on $v_j$ and the effect of $v_j$ on $v_i$ are different in edge $v_j \to v_i$, which produces two types of edges that represent two different transition relationships. One directed edge called outgoing edge with weights of $\omega_{ij}^{out}$ and the other directed edge called incoming edge with weights of ${\omega}_{ji}^{in}$. Their weights can be computed by:
\begin{align}
\omega_{ij}^\text{out} &= \frac{\operatorname{Count}\left(v_i, v_j\right)}{\sum_{v_i \to v_k}
	\operatorname{Count}\left(v_i, v_k\right)}, \\
{\omega}_{ij}^\text{in} &= \frac{\operatorname{Count}\left(v_j, v_i\right)}{\sum_{v_k \to v_i} \operatorname{Count}\left(v_k, v_i\right)},
\end{align}
where function $\operatorname{Count}(x, y)$ is used to calculate the number of occurrences that user interacts item $y$ after interacts item $x$. The topological structure of user behavior graph $\mathcal{G}^u$ can be represented by two adjacency matrices, which can be written as:
\begin{align}
\mathbf{A}^\text{out}_{u}[i,j] &= \omega_{ij}^\text{out},  \\
\mathbf{A}^\text{in}_{u}[i,j] &= {\omega}_{ij}^\text{in},	
\end{align}
$\mathbf{A}_u^{\text{out}}$ and $\mathbf{A}_u^{\text{in}}$  represent adjacency matrix of outgoing and incoming edges in user behavior graph.
\subsection{Personalized Graph Neural Network}
\label{sec:generate-item-user-embedding}

SR-GNN \cite{wu2019session} is the first model to apply GNNs in the session-based recommendation, which feeds the session graphs containing rich node connections into GGNN to automatically extract useful features of items. However, the GGNN used in SR-GNN is not suitable for personalized recommendation because it fails to inject the user's information into the graph model. To address this limitation, herein, we present personalized graph neural networks (PGNN), which is used to learn the complex item transition relationships between items interacted by each user, and then obtain the representation of items and users.

Various users have different behavior patterns, which results in different item transition relationships for each user. So, we consider the user factor when designing PGNN architecture. At each time of node update, we fuse user embedding $\mathbf{e}_u$ with the current representation of node $\mathbf{h}_i^{t-1}$. For example, at $t$ time, the aggregated incoming and outcoming information of node $i$ can be formulated as:
\begin{align}
\mathbf{a}_{\text{out}_i}^{(t)} &= \sum_{v_i \to v_j}
{\mathbf{A}^\text{out}_{u}[i,j]} \left[\mathbf{h}^{(t-1)}_j \parallel \mathbf{e}_u\right] \mathbf{W}_{\text{out}} \\
\mathbf{a}_{\text{in}_i}^{(t)} &= \sum_{v_j \to v_i }{\mathbf{A}^\text{in}_{u}[i,j]} \left[\mathbf{h}^{(t-1)}_j \parallel \mathbf{e}_u\right]\mathbf{W}_{\text{in}},\\
\mathbf{a}_i^{(t)} &=  \mathbf{a}_{\text{out}_i}^{(t)} \parallel  \mathbf{a}_{\text{in}_i}^{(t)},
\end{align}
where $\parallel$ is the concatenation operation. Because $\mathcal{G}^u$ is bidirectional, to embed bidirectional propagation information, we consider two parameters, $\mathbf{W}_{\text{in}}$ and $\mathbf{W}_{\text{out}}\in \mathbf{R}^{(d+d')\times \hat{d}} $, which transform the user and item connection vectors to two different $\hat{d}$-dimensional vectors, respectively. All users share parameters $\mathbf{W}_{\text{in}}$ and $\mathbf{W}_{\text{out}}$.

Then, we use gated recurrent units (GRUs) \cite{cho2014learning} to incorporate information from other nodes with hidden states of the previous timestep, and update each node's hidden state:
\begin{align}
\mathbf{z}^{(t)}_{i} & = \sigma\left(\mathbf{W}_z\mathbf{a}^{(t)}_{i}+\mathbf{U}_z\mathbf{h}^{(t-1)}_{i} \right), \label{eq:update-gate}\\
\mathbf{r}^{(t)}_{i} & = \sigma\left(\mathbf{W}_r\mathbf{a}^{(t)}_{i}+\mathbf{U}_r\mathbf{h}^{(t-1)}_{i} \right), \label{eq:reset-gate}\\
\widetilde{\mathbf{h}^{(t)}_{i}} & = \tanh\left(\mathbf{W}_o \mathbf{a}^{(t)}_{i}+\mathbf{U}_o \left(\mathbf{r}^{(t)}_{i} \odot \mathbf{h}^{(t-1)}_{i}\right)\right), \label{eq:candidate-state}\\
\mathbf{h}^{(t)}_{i} & = \left(1-\mathbf{z}^{(t)}_{i} \right) \odot \mathbf{h}^{(t-1)}_{i} + \mathbf{z}^{(t)}_{i} \odot \widetilde{\mathbf{h}^{(t)}_{i}}, \label{eq:final-state}
\end{align}
%
where $\mathbf{z}_i^t$ and $\mathbf{r}_i^t$ are update and reset gate, $\sigma(\cdot)$ is the sigmoid function, and $\odot$ is the element-wise multiplication operator. $\mathbf{W}_z$, $\mathbf{U}_z$, $\mathbf{W}_r$, $\mathbf{U}_r$, $\mathbf{W}_o$, $\mathbf{U}_o$ are GRU parameters shared by all users. After a total of $T$ propagation steps, the final hidden state vector $\mathbf{h}^{(T)}_i$ of each node $i$ can be obtained in graph $\mathcal{G}^u$. For convenience, we use $\mathbf{h}_i$ instead of $\mathbf{h}_i^{(T)}$. The final hidden state of each node not only contains its node features, but also aggregates the information from its $T$-order neighbors. 

Similar to most graph-based model \cite{wu2019session,qiu2019rethinking}, PGNN is suitable for the scenarios that the user repeatedly click the same items within sessions or across sessions. When the same user repeatedly clicks the same items across sessions, all sessions of the user can convert to a fully connected graph structure. Obviously, PGNN can capture the items transition pattern across sessions. For the other extreme case, the same user has no repeated interaction items across sessions. In this case, the user's behavior graph contains many disconnected sub-graphs, where each sub-graph corresponds to a session. Since all sessions of each user in PGNN share the same user embedding, each sub-graph in user's behavior graph can be related through the user embedding when the node embedding updates. Therefore, PGNN can still capture the association across sessions.

\subsection{Generating User's Unified Representation via Attention Networks}
\label{sec:generate-session-embedding}
In this section, we propose to use the Transformer's attention mechanism \cite{vaswani2017attention} to calculate the explicit effect of historical sessions on the current session and then get the dynamic representation of each user through the attention network. Finally, the user's unified representation can be obtained for personalized recommendation.
\subsubsection{\bf Calculating the impact of historical sessions on the current session}
\label{subsec:attention}
In our model, we resort to Transformer network \cite{vaswani2017attention} which is widely used in some popular neural machine translation models to complete the calculation of the impact of historical sessions on current session. The scaled dot-product attention mechanism is the core of Transformer network.

{\bf Transformer Attention}: The input of Transformer attention consists of queries and keys of dimension $d_k$, and values of dimension $d_v$. We compute the dot products of the {\it query} with all {\it keys} divide each by $\sqrt{d_k}$, then apply a softmax function to obtain the weight on values. The scaled dot-product attention is formally defined as:
\begin{align}
\operatorname{Attention}(\mathbf{Q}, \mathbf{K}, \mathbf{V}) = \operatorname{softmax}\left(\frac{\mathbf{Q}\mathbf{K}^\top}{\sqrt{d_k}}\right)\mathbf{V},
\label{eq:attenion}
\end{align}
where $\mathbf{Q}, \mathbf{K}, \mathbf{V}$ represent the queries, keys, and values respectively, and the scale factor $\sqrt{d}$ is to avoid exceedingly large dot products and speed up convergence.

The embedding representations of user $u$'s historical sessions $\mathcal{S}_\text{h}^u$ and current session $S_\text{c}^u$ can be obtained through the output of PGNN. The embedding vector of historical session $S_i^u = \{v_{i,1}, v_{i,2}, \dots, v_{i,{m_i}}\}$ in $\mathcal{S}_\text{h}^u$ can be represented as $\mathbf{f}_i^u \in \mathbb{R}^d$, which can be calculated by max-pooling:
\begin{align}
\mathbf{f}_{i,j}^u = \max_{1{\leq}j{\leq}d} \left(\mathbf{h}_{1,j},\mathbf{h}_{2,j},\dots, \mathbf{h}_{{m_i},j}\right).
\end{align}
Therefore, historical session sequence $\mathcal{S}_\text{h}^u =\{S_1^u, S_2^u, \dots, S_{n-1}^u\}$ can be expressed as an embedded matrix $\mathbf{F}^u = \left[\mathbf{f}_1^u, \mathbf{f}_2^u, \dots, \mathbf{f}_{n-1}^u \right]$. For current session $S_\text{c}^u$, we simply denote the embedding matrix as $\mathbf{H}^u=\left[\mathbf{h}_1, \mathbf{h}_2, \dots, \mathbf{h}_m\right]$. In our context, we use current session embedding to query historical session embedding, where the queries $\mathbf{Q}$ are determined by  $\mathbf{H}^u$, the keys and values are determined by $\mathbf{F}^u$. In specical, we project $\mathbf{F}^u$ and $\mathbf{H}^u$ to the same latent space through nonlinear transformation:
\begin{equation}
\begin{aligned}
\mathbf{Q}^u &= \operatorname{Relu}\left(\mathbf{H}^u \mathbf{W}^Q\right), \\
\mathbf{K}^u &= \operatorname{Relu}\left(\mathbf{F}^u \mathbf{W}^K\right), \\
\mathbf{V}^u &= \operatorname{Relu}\left(\mathbf{F}^u \mathbf{W}^V\right),
\end{aligned}
\end{equation}
where $\mathbf{W}^Q, \mathbf{W}^K, \mathbf{W}^V \in \mathbb{R}^{d \times d}$ are the projection matrices and shared by all users. The effect of the historical sessions on current session can be calculated by 
\begin{equation}
\mathbf{H}_h = \operatorname{Attention}\left(\mathbf{Q}^u,\mathbf{K}^u,\mathbf{V}^u\right).
\end{equation}
After the effect of history session on each item in the current session sequence is calculated, we then compute the embedding of the current session as follows:
\begin{equation}
\mathbf{H}^{u^\prime} = \mathbf{H}_h + \mathbf{H}^u.
\end{equation}
Then, the current session embedding $\mathbf{H}^{u^\prime}$ can be rewritten as $\left[\mathbf{h}_1^\prime, \mathbf{h}_2^\prime, \dots, \mathbf{h}_m^\prime \right]$.

\subsubsection{\bf Generating the user's unified representation} 
\label{subsec:user_rep}
The current session embedding $\mathbf{H}^{u^\prime}$ combines long- and short-term interests of users. In the following part, we describe how to encode $\mathbf{H}^{u^\prime}$ to the user unified representation vector for next-item recommendation task. 

Similar to SR-GNN \cite{wu2019session}, we first use the attention mechanism to encode the current embedding matrix to local representation and global representation, respectively, where local representation $\mathbf{z}_l$ denotes the user's recent interest and global representation $\mathbf{z}_g$ denotes the user's general interest.
$\mathbf{z}_l$ can be simply defined as $\mathbf{h}_m^\prime$, which is the embedding of last clicked item within the current session. It can be written as:
\begin{equation}
\mathbf{z}_l = \mathbf{h}_m^\prime.
\end{equation}
$\mathbf{z}_g$ is defined as:
\begin{align}
\mathbf{z}_g &= \sum_{i=1}^{m}\alpha_i \mathbf{h}_i^\prime, \\
\alpha_i &= \mathbf{W}_0\sigma\left(\mathbf{W}_1 \mathbf{h}_m^\prime + \mathbf{W}_2 \mathbf{h}_i^\prime + \mathbf{b}_c\right), 
\end{align}
where parameters $\mathbf{W}_0\in \mathbb{R}^d$, $\mathbf{W}_1, \mathbf{W}_2\in \mathbb{R}^{d\times d}$ control the weights of item embedding vectors, $\mathbf{b}_c \in \mathbb{R}^d$ is a bias vector, $\sigma(\cdot)$ denotes the sigmoid function and weighted coefficient $\alpha_i$ determines the weights of items of current session when making predictions.
After that, we compute the user's dynamic representation $\mathbf{z}_d$ as follows,
\begin{equation}
\mathbf{z}_d=\mathbf{z}_g \parallel \mathbf{z}_l.
\end{equation}

The embedding $\mathbf{e}_u$ implies the inherent attributes of the user and can be regarded as a static representation. So, we concatenate the dynamic and the static representation into one vector, then get the unified representation of users $\text{z}_u$ through linear transformation:
\begin{equation}
\text{z}_u=\mathbf{B} \left[\mathbf{z}_d \parallel \mathbf{e}_u\right],
\end{equation}
where matrix $\mathbf{B}\in \mathbb{R}^{d \times (2d + d^\prime)}$ compresses two combined embedding vectors into the latent space $\mathbb{R}^d$, and $d, d^\prime$ are the dimension of item and user embedding respectively.

\subsection{Making Recommendation}
\label{sec:make-recommendation}
After obtaining the unified representation of user $u$, we compute the recommendation score $\hat{\mathbf{z}}_i$ for each item $v_i\in V$.  The score function is defined as: 
\begin{equation}
\hat{\mathbf{z}}_i = {\mathbf{z}_u}^\top \mathbf{e}_{v_i},
\end{equation}
where $\mathbf{z}_u$ and $\mathbf{e}_{v_i}$ denote the user's unified representation and item $v_i$'s embedding, respectively.
Then we apply a $\operatorname{softmax}$ function to get the output vector:
\begin{equation}
\hat{\mathbf{y}} = \operatorname{softmax}(\hat{\mathbf{z}}),
\end{equation}
where $\hat{\mathbf{z}}\in{\mathbb{R}^{|V|}}$ denotes the recommendation scores over all candidate items $V$ and $\hat{\mathbf{y}}$ denotes the probabilities that items will be interacted by user $u$ in the next time of the current session $S_c^u$.

For any user behavior graph, the loss function is defined as the cross-entropy of the prediction and the ground truth. It can be written as follows,
\begin{equation}
\mathcal{L}(\hat{\mathbf{y}}) = -\sum_{i = 1}^{\left|V\right|}\mathbf{y}_i \log{(\hat{\mathbf{y}}_i)} + (1 - \mathbf{y}_i) \log{(1 - \hat{\mathbf{y}}_i)},
\end{equation}
where $\mathbf{y}$ denotes the one-hot encoding vector of the ground truth item. Finally, we use the back-propagation through time (BPTT) algorithm to train the proposed A-PGNN.

\section{Experiments}
We first describe the experimental setting from Section 4.1 to Section 4.4, and then compare A-PGNN against state-of-the-art methods in Section 4.5. To verify the effectiveness of two important components in our model,
we perform ablation studies in Section 4.6. In Section 4.7, we further give analysis about the effect of test session's characteristics on the model's performance. The hyper-parameter study is finally presented in Section 4.8. We intend to answer the following questions through experiments.

\begin{itemize}[leftmargin=*]
	\item {\bf RQ1}: How does A-PGNN perform compared with other SOTA models?
	\item{\bf RQ2}: What is the effect of various components in A-PGNN?
	\item{\bf RQ3}: How does A-PGNN  perform when dealing with test sessions with different lengths and sessions with different numbers of historical sessions? 
	\item{\bf RQ4}: What are the effects of different hyper-parameter settings (parameter initialization methods, maximum historical session and PGNN propagation step) on A-PGNN? 
\end{itemize}
\begin{table}%
	\caption{Statistics of data sets after preprocessing}
	\label{tab:one}
	\begin{minipage}{\columnwidth}
		\begin{center}
			\resizebox{0.8\columnwidth}{!}{
				\begin{tabular}{lll}
					\toprule
					{\bf Dataset}   & {\bf Xing}& {\bf Reddit}\\
					\hline\hline
					users           & 11479     &18271\\
					items           & 59121     &27452\\
					Sessions        & 91683     &1135488\\
					Average session length      & 5.78  &3.02\\
					Sessions per user   & 7.99  &62.15\\
					Train sessions   & 69135      &901161\\
					Test sessions   & 22548      &234327\\
					\toprule
			\end{tabular}}
		\end{center}
	\end{minipage}
\end{table}%
\subsection{Data Sets}
\label{dataset}
We used two different real-word data sets for our experiments. The first is the Xing data \cite{quadrana2017personalizing}, which is released from RecSys Challenge 2016\footnote{\url{http://2016.recsyschallenge.com/}}. The second is a data set \cite{Ruocco2017Inter} extracted from the social news and discussion website Reddit\footnote{\url{https://www.kaggle.com/colemaclean/subreddit-interactions}}.

{\bf Xing}. The Xing data set contains interactions on job postings for 770,000 users over an 80-day period. In these data, user behaviors include click, bookmark, reply, and delete. Following the preprocessing procedure of \cite{quadrana2017personalizing}: We split the Xing data into session by 30-minute idle threshold and discarded interactions having typed "delete."  Also discarded  are repeated interactions of the same type within sessions to reduce noise (e.g. repeated clicks on the same job posting within a session). Removed sessions having less than 3 interactions to filter too short and poorly informative sessions, and kept users having 5 sessions or more to have sufficient cross-session information.

{\bf Reddit}. The Reddit data set contains tuples of user name, a subreddit where the user makes a comment to a thread, and a timestamp for the interaction. We split each user’s records into sessions manually by using the same approach as mentioned in \cite{Ruocco2017Inter}. The time threshold turn to be 60-minutes this time.

Then we preprocessed both data sets as follows: For each user, we hold the first $80\%$ of his sessions as the training set. The remaining $20\%$ constitutes the test set. We tune the hyper-parameters of the algorithms on the last 10\% of the training set. The statistics of two data sets after the preprocessing steps are shown in Table \ref{tab:one}. Referring to \cite{wu2019session}, we segment each user's sessions $\mathcal{S}^u$ into a series of sequences and labels. For example, for an input $\mathcal{S}^u=\{\{v_{1,1},v_{1,2},v_{1,3}\},\{v_{2,1},v_{2,2}\},\{v_{3,1},v_{3,2},v_{3,3}\}\}$ of user $u$, where $\mathcal{S}_\text{h}^u =\{\{v_{1,1},v_{1,2},v_{1,3}\},\{v_{2,1},v_{2,2}\}\}$, $\mathcal{S}_\text{c}^u = \{\{v_{3,1},v_{3,2},v_{3,3}\}\}$. We generate historical sessions, current sessions, and labels,
$\mathcal{S}_{\text{h}_1}^u = \{\{v_{1,1},v_{1,2},v_{1,3}\}\}$, 
$\mathcal{S}_{\text{c}_1}^u = \{\{v_{2,1}\}\}$, $\text{label}_1= v_{2,2}$; 
$\mathcal{S}_{\text{h}_2}^u = \{\{v_{1,1},v_{1,2},v_{1,3}\},\{v_{2,1},v_{2,2}\}\}$, $\mathcal{S}_{\text{c}_2}^u = \{\{v_{3,1}\}\}$, $\text{label}_2= v_{3,2}$; $\mathcal{S}_{\text{h}_3}^u  = \{\{v_{1,1},v_{1,2},v_{1,3}\},\{v_{2,1},v_{2,2}\}\}$,  $\mathcal{S}_{\text{c}_3}^u =\{\{v_{3,1},v_{3,2}\}\}$, $\text{label}_3=v_{3,3}$, where the label is the next interacted item within the current session. 

\begin{table*}
	\centering
	\caption{Performance of A-PGNN and nine compared models in terms of Recall@5, 10, 20, and Mrr@5, 10, 20 on two data sets. The * and underlined numbers mean the best results on traditional and deep neural methods, respectively. $Improvement^-$ means improvement over the best conventional methods. $Improvement^*$ means improvement over the best deep neural methods.}
	\setlength{\tabcolsep}{1mm}
	\resizebox{2.05\columnwidth}{!}{
		\begin{tabular}{cccccccccccccc|}
			\toprule
			\multirow{2}[0]{*}{\bf Data} & \multicolumn{6}{c}{\bf Xing} & \multicolumn{6}{c}{\bf Reddit} \\ \cmidrule[0.5pt](lr){2-7} \cmidrule[0.5pt](lr){8-13}
			&\bf Recall@5 & \bf Recall@10 & \bf Recall@20 & \bf Mrr@5 &\bf Mrr@10 &\bf Mrr@20 & \bf Recall@5 & \bf Recall@10 &\bf Recall@20 &\bf Mrr@5 &\bf Mrr@10 &\bf Mrr@20  \\ \midrule
			Pop           & 0.21   &0.26  &0.58  &0.08 &0.09 &0.11  &13.22  &19.46   &26.47 &8.50  &9.32  &9.82\\
			Item-KNN      & 8.79   &11.85 &14.67 &5.01 &5.42 &5.62  &21.71  &30.32   &38.85 &11.74 &12.88 &13.49\\
			FPMC          & 1.70   &2.42  &3.27  &0.61 &0.50 &0.37  &29.91  &34.31   &44.32 &8.78  &6.56  &4.54\\
			SKNN    &14.36 & 19.42 & 24.12 &9.29 &9.8 &10.22 &\underline{34.29}   &42.17 &\underline{49.68} &\underline{19.11} &\underline{20.16} &\underline{20.68}\\
			VSKNN    &\underline{14.46} &\underline{19.60} &  \underline{24.25} &\underline{9.48} &\underline{10.07}&\underline{10.39}&34.25   &\underline{42.17} &49.67 &19.09 &20.14 &20.67\\ \midrule
			GRU4Rec       & 10.35   &13.15  &15.30 &5.94 &6.36 &6.69  &33.72	&41.73	 &50.04	&24.36 &25.42	&26.00\\
			
			SR-GNN         &13.38   &16.71*  &19.25& 8.95& 9.39&9.64 &34.96  &42.38   &50.33 &25.90 &26.88   &27.44\\
			H-RNN         & 10.74  &14.36 &17.64 &7.22 &7.78  &8.83  &{44.76}	&{53.44}	 &{61.80}	&{32.13} &{33.29}	&{33.88}\\
			HierTCN        & 13.57*  &16.55 &19.93* &9.23* &9.48*  &10.23*  &{47.15}*	&{55.37}*	 &{63.96}*	&{32.18}* &{33.79}*	&{34.27}*\\ \midrule
			\bf A-PGNN  & \bf 14.38 &\bf{17.06}&\bf{19.98}&\bf10.36&\bf10.71 &\bf10.91 &\bf49.19	&\bf59.43&\bf68.00&\bf33.54&\bf34.92&\bf35.52\\
			
			$Improvement^-$ &- &-&-&9.28\%&6.36\%&5.00\%&43.45\%&40.93\%&36.88\%&85.96\%&73.21\%&71.76\%\\
			$Improvement*$ &5.97\% &2.09\%&0.25\%&12.24\%&12.97\%&6.65\%&4.33\%&7.33\%&6.32\%&4.23\%&3.34\%&3.65\%\\
			\bottomrule
		\end{tabular}
	}
	\label{tab:result-baseline}
\end{table*}
\subsection{Compared Methods}
We compared the performance of our proposed A-PGNN with nine compared methods, including conventional methods and deep neural methods.
\begin{itemize}[leftmargin=*]
	
	\item {\bf POP} recommends the top $K$ frequent items in the training set.
	
	\item {\bf Item-KNN}\cite{Sarwar2001} computes an item-to-item cosine similarity based on the co-occurrence of items within sessions.
	
	\item {\bf FPMC}\cite{rendle2010factorizing} is a sequential prediction method based on the personalized Markov chain.
	
	\item {\bf SKNN}\cite{jannach2017recurrent} selects the $K$ most similar sessions from the training set to retrieve candidate items for recommendation
	\item {\bf VSKNN}\cite{ludewig2018evaluation} is a sequential extension based SKNN.
	
	\item {\bf GRU4Rec}\cite{tan2016improved} applies improved RNNs in session-based recommendation scenario.
	\item {\bf SR-GNN} \cite{wu2019session} utilizes the Gated Graph Neural Networks to capture the complex transition relationships of items for the session-based recommendation.
	
	\item {\bf H-RNN} \cite{quadrana2017personalizing}\cite{Ruocco2017Inter}  use a hierarchical RNNs consist of a session-based and a user-level RNN to model the cross-session evolution of the user's interest. Due  to \cite{quadrana2017personalizing} and \cite{Ruocco2017Inter} are similar in model architecture, we only select the best results of the two models as a comparison, collectively referred to as H-RNN.
	
	\item {\bf HierTCN} \cite{you2019hierarchical} utilizes the hierarchical architecture that contains RNN and Temporal Convolutional Network to capture both the long-term interests and short-term interactions. 
	
\end{itemize}
\subsection{Evaluation Metrics}
Following the metrics are used to evaluate each methods, which are also widely used in other related works \cite{quadrana2017personalizing, Ruocco2017Inter}.

{\bf Recall@K} (Precision) is widely used as a measure for predictive accuracy. It represents the proportion of correctly recommended items among the top-$K$ items.

{\bf MRR@K} (Mean Reciprocal Rank) is the average of reciprocal ranks of the correctly-recommended items. When all the rank positions exceed $K$, the reciprocal rank is set to 0. The MRR measure considers the order of recommendation ranking, where large MRR value indicates that correct recommendations are at the top of the ranking list.

We used Recall@K and MRR@K with $K = 5, 10, 20$ to evaluate all compared methods.

\subsection{Parameter Setup}
We set the dimension of item embedding $d=100$ for Xing as \cite{quadrana2017personalizing}, $d=50$ for Reddit as \cite{Ruocco2017Inter}, and set user embedding dimension $d'=50$ for both data sets. According to the data processing method of Section \ref{dataset}, the maximum length of current session is 20. Because of the limitation of computing resources, for each user, we limit the number of historical sessions that feed into our model, that is, only his $M$ most recent historical sessions can be utilized to assist with making prediction for current session. We set the $M$ to a be a hyper-parameter named as "maximum historical session". For Xing, M is 50, whereas for Reddit, it is 30. As for the PGNN's propagation step $T$, we set $T$ to be 1 for Xing and 3 for Reddit. All parameters are initialized by using Uniform distribution $\mathcal{U}(-1/\sqrt{d},1/\sqrt{d})$. The model is trained with Adam \cite{Kingma2014Adam} optimizer, with learning rate 0.001. The coefficient of L2 normalization is set to 0, and the batch size is 100. In particular, we use batch normalization \cite{ioffe2015batch} between dot-attention layer and session-attention layer to prevent from overfitting on smaller Xing data. For the baseline methods, we use the default hyperparameters except for dimensions. We run the evaluation 5 times with different random seeds and report the mean value per algorithm.

\begin{table*}
	\centering
	\caption{Performance of A-PGNN compared with four ablation models in terms of Recall@5,10 and Mrr@5,10 on two data sets. The numbers in parentheses indicate the percentage of performance degradation of the ablation model compared to A-PGNN.}
	\setlength{\tabcolsep}{1mm}
	\resizebox{2.05\columnwidth}{!}{
		\begin{tabular}{cccccccccccccc}
			\toprule
			\multirow{2}[0]{*}{\bf Datasets} & \multicolumn{4}{c}{\bf Xing} & \multicolumn{4}{c}{\bf Reddit} \\ \cmidrule[0.5pt](lr){2-5} \cmidrule[0.5pt](lr){6-9}
			&\bf Recall@5 & \bf Recall@10  & \bf Mrr@5 &\bf Mrr@10  & \bf Recall@5 & \bf Recall@10  &\bf Mrr@5 &\bf Mrr@10 \\ \midrule
			A-PGNN(\bf -U)  & 14.20(-1.25\%)	&16.69(-2.16\%)		&10.29(-0.68\%)		&10.81(-0.91\%) &48.97(-0.46\%)	&59.24(-0.32\%)		&33.37(-0.51\%)	&34.75(-0.49\%)	\\
			A-PGNN(\bf -A)   & 11.89(-17.3\%)	&14.74(-13.6\%)		&8.08(-2.21\%)	&8.46(-21.0\%) &49.04(-0.32\%)	&59.31(-0.20\%)   &33.35(-0.56\%)  &34.73(-0.54\%)	\\
			A-PGNN(\bf -P)  &13.84(-13.5\%)	&16.67(-2.28\%)		&9.67(-6.71\%)	&10.05(-6.16\%)	 &48.77(-0.87\%)  &58.75(-1.14\%)    &33.46(-0.24\%) &34.79(-0.37\%)  \\
			A-PGNN(\bf-A-P) &13.15(-8.55\%)   &16.29(4.51\%) 	&8.63(-16.7\%)	&9.33(-12.89\%) &32.60(-33.7\%) &40.06(-32.6\%) &23.09(-31.2\%)	&24.09(-31.0\%)  \\
			\midrule
			\bf{A-PGNN}     & \bf 14.38 &\bf17.06 &\bf10.36&\bf10.71  &{\bf 49.20}	&{\bf{59.43}}  &\bf{33.54}  &\bf{34.92}\\
			\bottomrule
		\end{tabular}
	}
	\label{tab:result-alation}
\end{table*}

\subsection{Performance Comparison (RQ1)}
\label{performance comparison}

First, for question {\bf RQ1}, we compare it with other state-of-the-art personalized and pure session-based recommendation methods. Table \ref{tab:result-baseline} reports the performance comparison results. We have the following observations:

Compared to "pure" session-based methods SR-GNN and GRU4Rec, the performance of A-PGNN and HierTCN verify that incorporating historical session information can improve the recommendation ability. HierTCN performs better than the H-RNN model on both data sets, indicating that Temporal Convolution Network has a more powerful sequence encoding capability than that of the GRU net. SR-GNN outperforms GRU4Rec and H-RNN by large margins on Xing, which attributes to the superiority of the graph-based model. In particular, prior study\cite{quadrana2017personalizing} on Xing has shown that users' activity within and across sessions has a high degree of repetitiveness, which means their behaviors are more easily to form graph structure. Therefore, the graph-based methods are more effective \cite{qiu2019rethinking}, \cite{wu2019session}.


It is obvious that deep neural methods perform better than conventional methods in most cases. However, the non-neural model VSKNN and SKNN exhibit strong competitive performance in Xing, which is better than most neural models. A possible explanation is that, in a job-seeking website, users are interested in the same kind of jobs, and their interacted items among sessions are quite similar. So sequence knn model produced excellent results by retrieving items in the $K$ most similar past sessions in the training data. However, A-PGNN still outperforms VSKNN by 9.28\%, 6.36\%, 5.00\% in Mrr@5/10/20. In contrast, VSKNN and SKNN perform poorly on Reddit than neural models such as A-PGNN, HierTCN, and HRNN,  especially in terms of Mrr values. One possible reason is that the average session length of Reddit is relatively smaller than that of Xing. Regarding entertainment, social, and news websites, user activity within-session has a high degree of variability, making it difficult to extract useful information from similar sessions only.

A-PGNN consistently yields the best performance on all the data sets compared with the state-of-the-art session-aware method HierTCN. We attribute the success of A-PGNN and HierTCN to their ability to model the effect of the user’s historical interests on the current session. However, they act explicitly or implicitly. As for HierTCN, it fuses the representations of all historical sessions into a single vector to represent the user's long term interest, which limits the effective use of historical information. In contrast, A-PGNN overcomes this deficiency by utilizing the dot-product attention mechanism to explicitly calculate the impact of historical sessions on the current session, which makes it better to integrate long-term and short-term preferences of users.
\subsection{Ablation Study (RQ2)}
\label{ablation study}

Next, turn to {\bf RQ2}, we compare our method with different variants to verify the effectiveness of two critical components, the Dot-Product Attention mechanism and PGNN.
{\bf A-PGNN(-U)}: APGNN without using user embedding;
{\bf A-PGNN(-A)}: A-PGNN without the Dot-Attention mechanism, that is, it does not consider the explicit impact of historical sessions on the current session;
{\bf A-PGNN(-P)}: A-PGNN has no PGNN component;
{\bf A-PGNN(-A-P)}: A-PGNN neither has the PGNN component nor the Dot-Product attention mechanism, which is equivalent to the session-based method.
We show the Reall@5/10, Mrr@5/10 results in Table \ref{tab:result-alation}, and have the following findings.


A-PGNN is consistently superior to A-PGNN(-P) and A-PGNN(-A). It illustrates the importance of personalized Graph Neural Network and explicit modeling of historical information. The A-PGNN performs relatively better than A-PGNN(-U), which suggests that PGNN is more capable of capturing relationships between items than vanilla GNN. 

A-PGNN(-A-P) outperforms A-PGNN(-A) on Xing. One possible reason is that directly combining historical sessions with the current session to construct user behavior graph may bring noise to the prediction of current session, and does not necessarily lead to improvement. A-PGNN performs better than A-PGNN(-A-P) and A-PGNN(-P), this might be that the PGNN and Dot-Product attention mechanism can mutually reinforce each other: PGNN could be used to capture the complex transition between items,  while Dot-Product attention can distinguish important historical session information.

A-PGNN(-A) and A-PGNN(-P) improve by more than 30\% compared to A-PGNN(-A-P) on Reddit. It again verifies the significance of Dot-Product attention and PGNN but they use different ways to utilize historical session information. Dot-Product attention explicitly calculates the impact of historical sessions on the current session, and PGNN can capture the complex item transition in each user's sessions in user special.

\subsection{The Effects of Current Session Length and the Number of Historical Sessions (RQ3)}
To answer {\bf RQ3}, we further analyze the effects of current session length and the numbers of historical sessions on performance. 
\begin{figure*}[t]
	\centering
	\subfloat[Recall@5 on Xing]{
		\begin{minipage}[t]{0.25\linewidth}
			\includegraphics[scale=0.5]{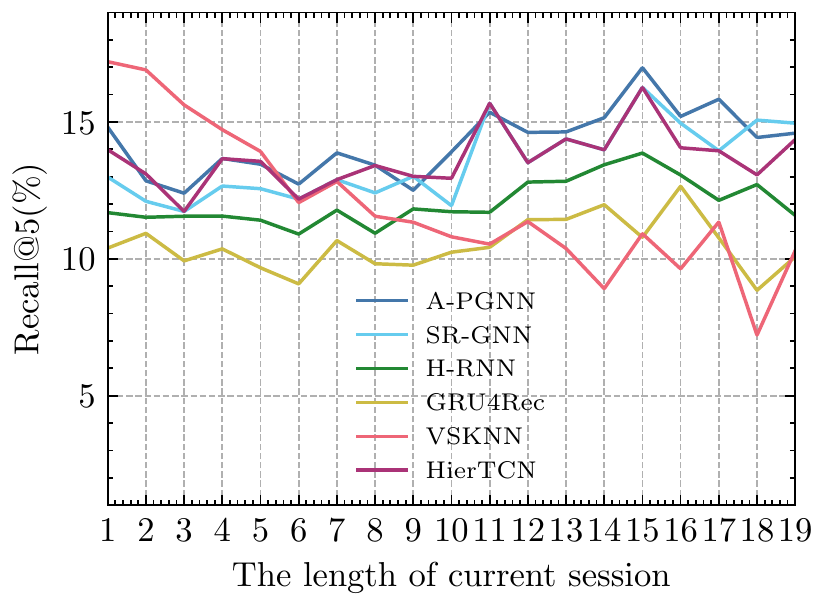}
			\label{fig_first_case}
		\end{minipage}
	}
	\subfloat[Mrr@5 on Xing]{
		\begin{minipage}[t]{0.25\linewidth}
			\includegraphics[scale=0.5]{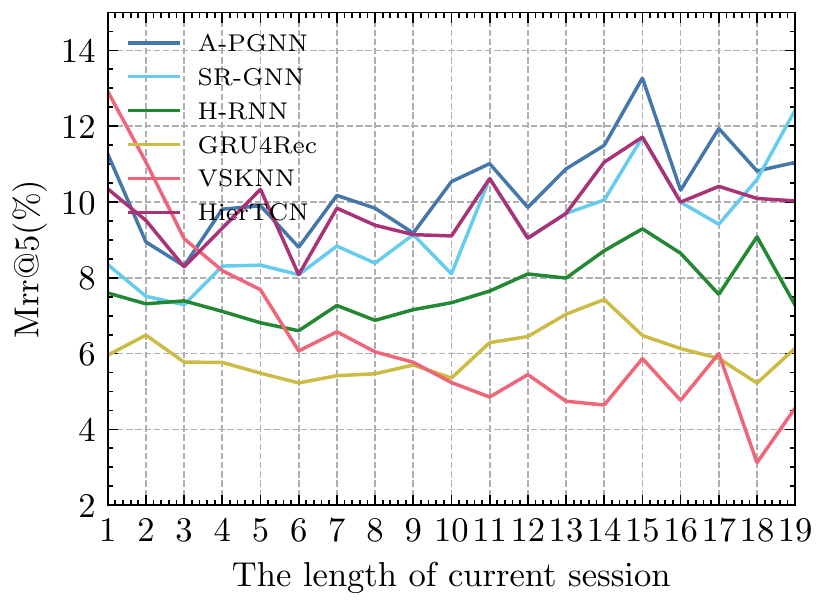}
			\label{fig_second_case}
		\end{minipage}
	}
	\subfloat[Recall@5 on Reddit]{
		\begin{minipage}[t]{0.25\linewidth}
			\includegraphics[scale=0.5]{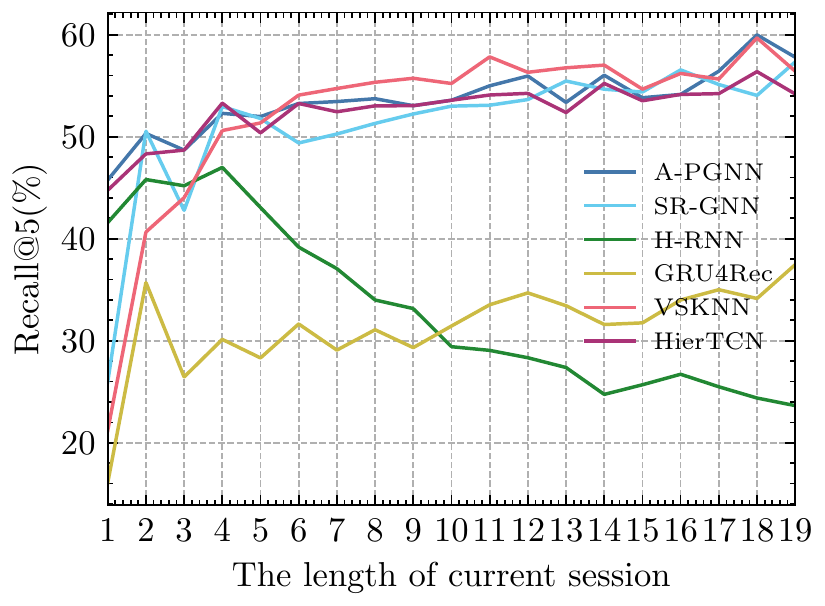}
			\label{fig_first_case}
		\end{minipage}
	}
	\subfloat[Mrr@5 on Reddit]{
		\begin{minipage}[t]{0.25\linewidth}
			\includegraphics[scale=0.5]{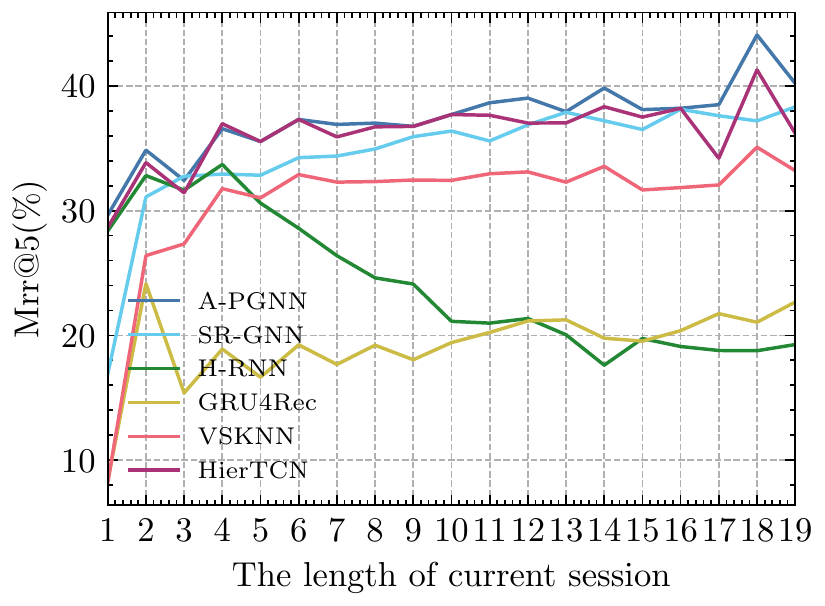}
			\label{fig_second_case}
		\end{minipage}
	}
	\centering
	\caption{Performance comparison in terms of Recall@5 and Mrr@5 tested on current sessions with different length on Xing data(a)(b) and Reddit data(c)(d).}
	\label{position_session}
\end{figure*}


\begin{figure}[th]
	\centering
	\subfloat[Xing data]{\includegraphics[scale=0.5]{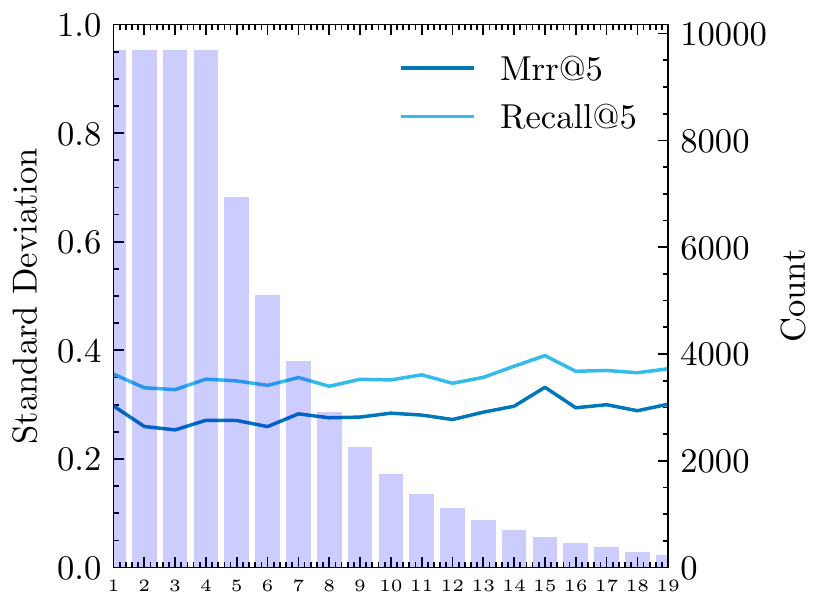}
		\label{var_session1}}
	\subfloat[Reddit data]{\includegraphics[scale=0.5]{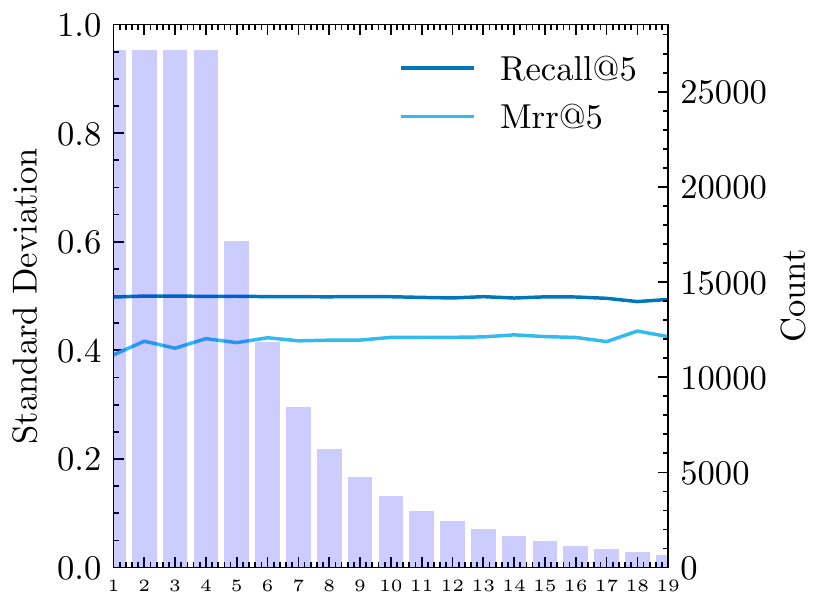}
		\label{var_session2}}
	\caption{The histogram shows the counts of sessions with different length and the line chart shows the standard deviation of indicator tested on sessions with different length.}
	\label{curr_session}
\end{figure}

\begin{figure}[th]
	\centering
	\subfloat[Xing data]{\includegraphics[scale=0.5]{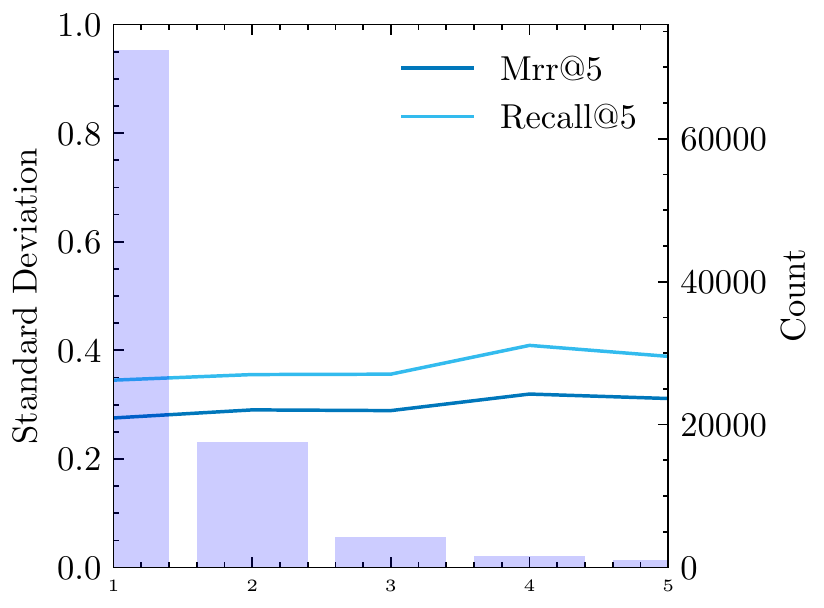}
		\label{var_session1}}
	\subfloat[Reddit data]{\includegraphics[scale=0.5]{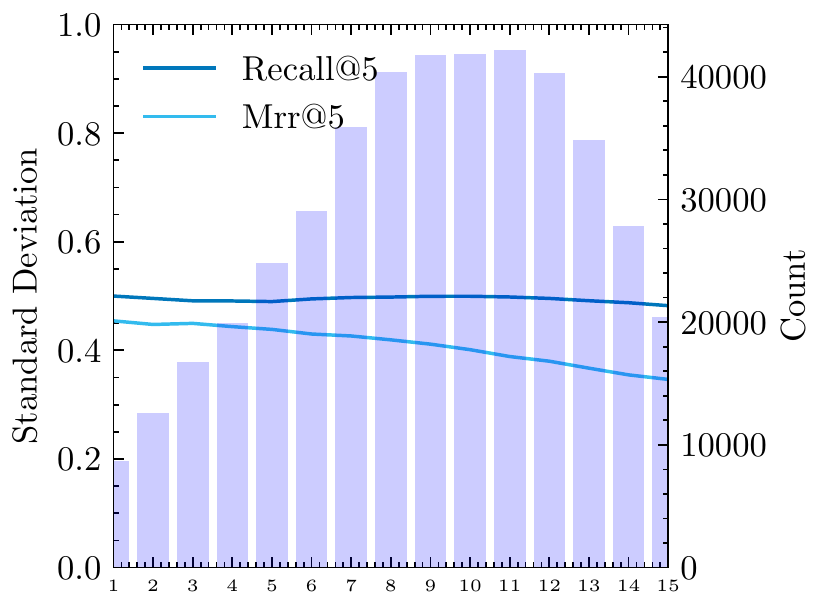}
		\label{var_session2}}
	\caption{The histogram shows the counts of test cases within each group and the line chart shows the standard deviation of indicator tested within each group.}
	\label{his_session}
\end{figure}

\begin{figure}[!t]
	\centering
	\includegraphics[width=3.2in]{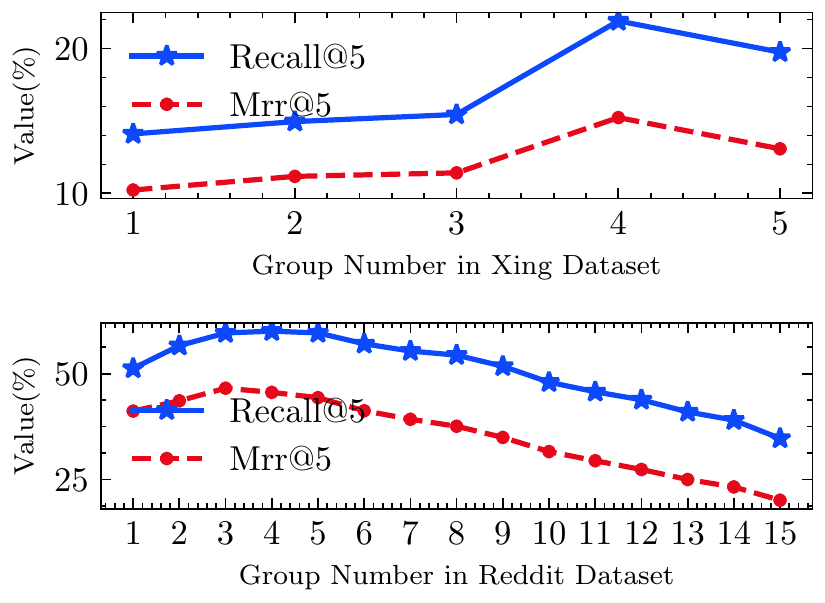}
	\caption{Performance of A-PGNN in terms of Recall@5 and Mrr@5 with different numbers of historical sessions.  }
	\label{historical_session}
\end{figure}

\subsubsection{The effect of current session length}
\label{session_length}
First, we analyze the capability of different models to handle  current sessions with different lengths. Similar to H-RNN\cite{quadrana2017personalizing}, we limit the analysis to sessions having length $\geq 5$ (8751 sessions for Xing and 26980 for Reddit). The histograms in Figure \ref{curr_session} show the counts of test cases under different length of current session. For comparison, we evaluate the recommendation performance of A-PGNN with HierTCN, H-RNN, SR-GNN, GRU4Rec, and VSKNN on different length of current sessions, respectively. Figure \ref{position_session} shows the Recall@5 and Mrr@5 results. Figure \ref{curr_session} line charts show the standard deviation (STD) of A-PGNN results under each length of current sessions. Because the maximum length of session is 20, the length of current session ranges from 1 to 19. From these results, some interesting conclusions can be drawn.

On the Xing data set, A-PGNN consistently outperforms other neural models on almost all length sessions. It outperforms other methods in terms of Recall@5 and Mrr@5 when the length is greater than 6 and 3, respectively. In this case, A-PGNN and HierTCN have an advantage over SR-GNN in most length sessions. H-RNN also outperforms GRU4Rec on all sessions. This shows the superiority of personalized models over "pure" session-based models. It is worth noting that the non-neural model VSKNN has achieved surprisingly good results in session length from 1 to 3. This may be because the users' interests are more concentrated in the first few clicks. Therefore, VSKNN can easily produce good results by looking for similar sessions to recommend items. However, the limitations of its model capabilities make it difficult to capture the evolution of interests, so the performance deteriorates as the session length increases. 

On the Reddit data set, A-PGNN outperforms SR-GNN in recommendation accuracy by a large margin (up to 43\% Recall@5 and 43\% Mrr@5) on sessions with length 1. With the increase of current session length, SR-GNN also becomes more competitive. When it comes to RNN-based models H-RNN and GRU4Rec, the advantage of H-RNN in short sequences is especially apparent. Nevertheless, as the length of the current session increases, we find that its performance in terms of Recall@5 and Mrr@5 becomes worse than GRU4Rec when session length greater than 10 and 12, respectively. A possible explanation is that, for Reddit, user's clicks at the beginning of the current session depend on historical user interactions. As the session becomes longer, user interest drifts gradually. In this case, the evolution of user interest mainly depends on the current session. Overuse of historical information may bring some invalid or interference information, which makes H-RNN hard to deal with sessions with larger length. In comparison, our model maintains stability in this case. Thanks to the attention mechanism that could explicitly model the impact of historical sessions on the current session, which reduces the effects of interference sessions. The mechanism of HierTCN dynamically updating items might alleviate this drawback, but it still underperforms A-PGNN in long sessions. On the Reddit data set, VSKNN performs worse on short sessions than on long sessions. Our analysis believes that the interests of users are diverse in each session, and it needs to take a more extended session to extract their interest. Therefore, VSKNN performance continues to increase and gradually stabilizes as the length of the session increases.

From Figure \ref{curr_session}, we find that for both data sets, the counts of current session with different length obey long-tailed distributions. However, the STD of the performance tested under each length has little variation, which shows that A-PGNN's performance is relatively stable.

\subsubsection{The effect of the number of historical sessions} 
When making predictions within the current session, is it the more historical sessions we incorporate, the better the recommendation performance will be?
To answer this question, for each test session, we group the test sessions by the number of historical sessions they own, which is denoted as $H$. For Xing, $H \in[1, 50]$ whereas for Reddit, $H \in[1, 150]$. To facilitate the analysis of indicator change trends, we partition the test sessions into several groups by units of ten. For xing, the testing sessions can be divided into 5 groups $[1, 10)$, $[10, 20)$, ..., $[40, 50]$. Also, Reddit can be divided into 15 groups $[1, 10)$, $[10, 20)$, ...., $[140, 150]$. Figure \ref{his_session} shows the counts of test cases within each group and the STD of indicator tested within each group.

The experiment results are shown in Figure \ref{historical_session}. For Xing, testing sessions in group 4 and 5 generally get a higher performance compared with those in group from 1 to 3. It suggests that those testing sessions with a larger amounts of historical sessions to assist with prediction generally perform better. For Reddit, the performance rises to a peak at group 3 then continues to fall for the rest of the period. A-PGNN has achieved satisfactory results in group from 1 to 8, while the performances of group from 9 to 15 are even worse than group 1. This shows that as the amount of historical session continues to increase, that is, after approximately greater than 100, the effectiveness of the model begins to deteriorate with the increase of user history length. This disproves the assumption that "more is better." From the aspect of reality, Xing is an employment-oriented website, which means that user interest drift is small and action in current session is strongly correlated to historical sessions. In contrast, Reddit is an entertainment, social , and news website, the purpose of the user's browsing behavior is often unclear, and it is susceptible to drift due to the content posted on the website. So, the interest of a long period in the past may bring noise to the predication of the current test session. In summary, we could choose to retain the appropriate number of historical sessions in actual scenarios according to the characteristic of the data set.

What is more, through Figure \ref{his_session}, it is found that the number of historical sessions owned by each test session is unevenly distributed, especially Xing, which follows the long-tail distribution. In Xing data set, the STD of the group with a small number of sessions is slightly bigger than that of group with a larger number of session. In Reddit data set, most test cases are mainly concentrated in the 7th to 13th groups. The STD of Recall@5 varies little for cross-groups. From the results of groups 11 to 15, we can see that, as the counts of test cases within each group continue to drop, the STD of Mrr@5 continues to get smaller. 



\subsection{Hyper-Parameter Study (RQ4)}
\label{hyperparameter}
We first conduct a sensitivity analysis of the model parameter initialization. Then, we perform experiments to explore how the hyperparameters like maximum historical sessions and PGNN propagation steps influence the performance. 

\subsubsection{Parameter sensitivity analysis}
We evaluate the performances of A-PGNN with a variety of choices of initialization mechanisms, including Gaussian $\mathcal{N}(0,0.1)$, Uniform $\mathcal{U}(-1/\sqrt{d}, 1/\sqrt{d})$, Truncated Gaussian $\mathcal{N}(0, 0.1)$, and Xavier initialization\cite{glorot2010understanding}. Figure \ref{sensitivity} shows the experiment results obtained with both data  sets. It can be observed that the performance of A-PGNN on Xing with different initialization methods varies greatly and achieves the best results under Uniform initialization, while the results on Reddit data set are more stable. So A-PGNN is more sensitive to parameter initialization on Xing than on Reddit. The reason might be that the small scale of Xing data causes the model to converge quickly, which makes the model more sensitive to the initialization method on this data set, whereas it is more stable on a large scale data set, Reddit.

\begin{figure}[ht]
	\centering
	\subfloat[Xing data]{\includegraphics[scale=0.5]{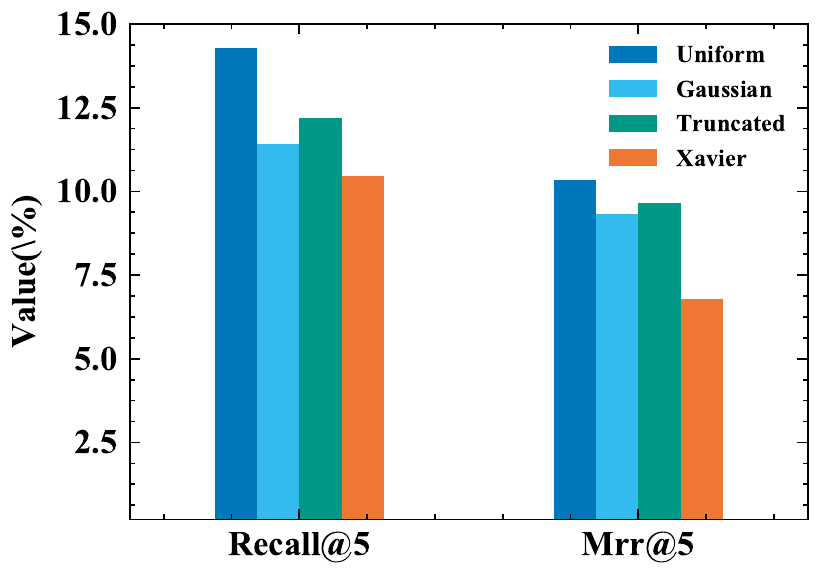}
		\label{fig_first_case}}
	\subfloat[Reddit data]{\includegraphics[scale=0.5]{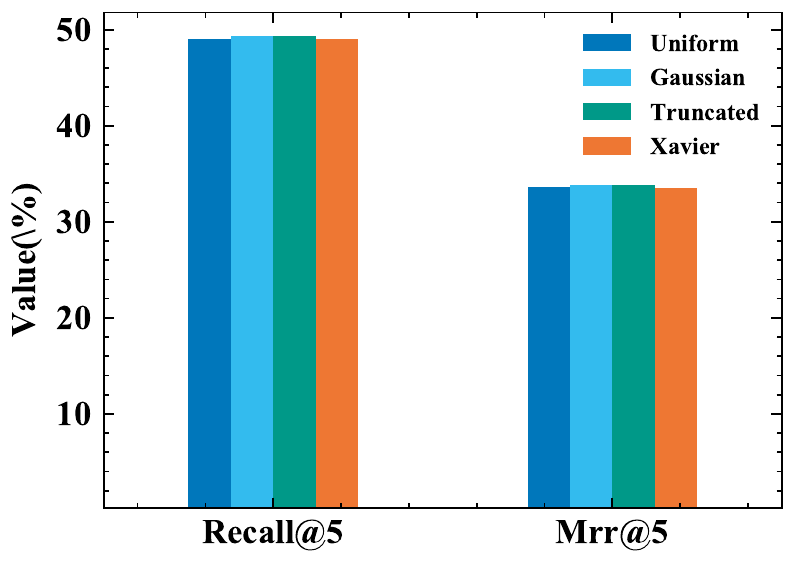}
		\label{fig_second_case}}
	\caption{
		Performance of A-PGNN in terms of Recall@5 and MRR@5 with different parameter initialization methods. \
	}
	\label{sensitivity}
\end{figure}

\subsubsection{Effect of maximum historical session}
\label{max_historical_session}
In this subsection, we investigate how the performance change with hyperparameter maximum historical session $M$, which indicates the upper limits of historical information that the network can utilize to make predictions for current session. Figure \ref{maxsession} shows the evaluation values with maximum historical sessions $M$. Recall@5 reaches its highest level when M is 40 for Xing and M is 3 for Reddit. And it then continues to fall. Although the same thing does not happen on Xing’s performance of Mrr@5, we can make a general conclusion that higher $M$ does not lead to better results. This shows that the increase in user historical information does not necessarily lead to an increase in model performance.

\subsubsection{Effect of PGNN propagation steps}
As the PGNN is a pivotal component of our model, we investigate the effect of propagation step $T$ mentioned in Section \ref{sec:generate-item-user-embedding}. What is worth mentioning is that the propagation steps mentioned in GGNN\cite{DBLP:journals/corr/LiTBZ15} is consistent with the Graph Neural Network layers. 

The experimental results are shown in Figure \ref{pgnn_layer}. We can see that, for Xing and Reddit, the performance reaches the highest value when $T$ is 1 and 3, respectively, and then gradually deteriorates with the increase of $T$. The value of the highest point is great than the value of T=0, which also verifies the effectiveness of PGNN. From the perspective of data type, for users in Xing, their interactive items tend to be of similar categories or topics. For example, they tend to look through the same kind of job postings for a specific career, which may suggest that node embeddings in the same user behavior graph are more likely to be closer in the embedding space compared with that of Reddit. The use of far propagation steps in Xing may lead to over-smoothing. In summary, it is more reasonable to choose a smaller T for Xing, and a larger $T$ for Reddit.

\begin{figure}[t]
	\centering
	\subfloat[Xing data]{\includegraphics[scale=0.5]{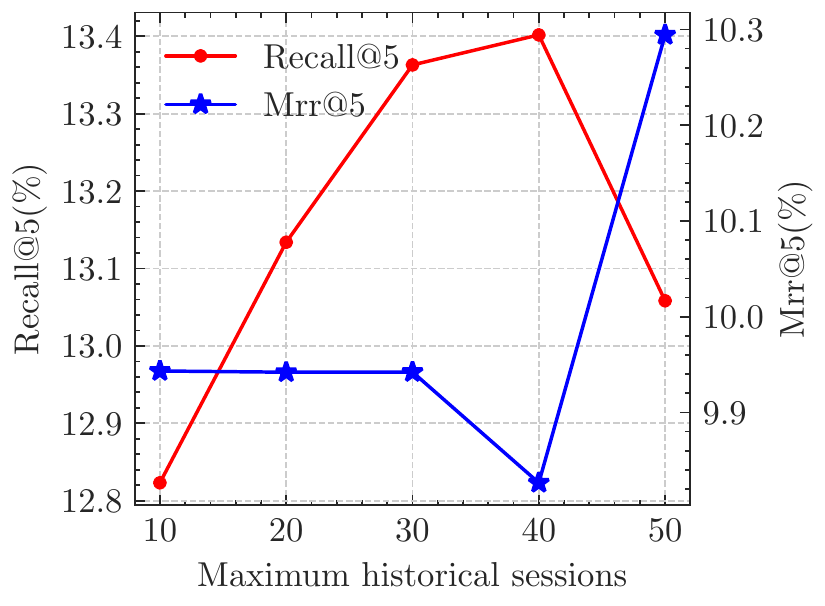}
		\label{fig_first_case}}
	\subfloat[Reddit data]{\includegraphics[scale=0.5]{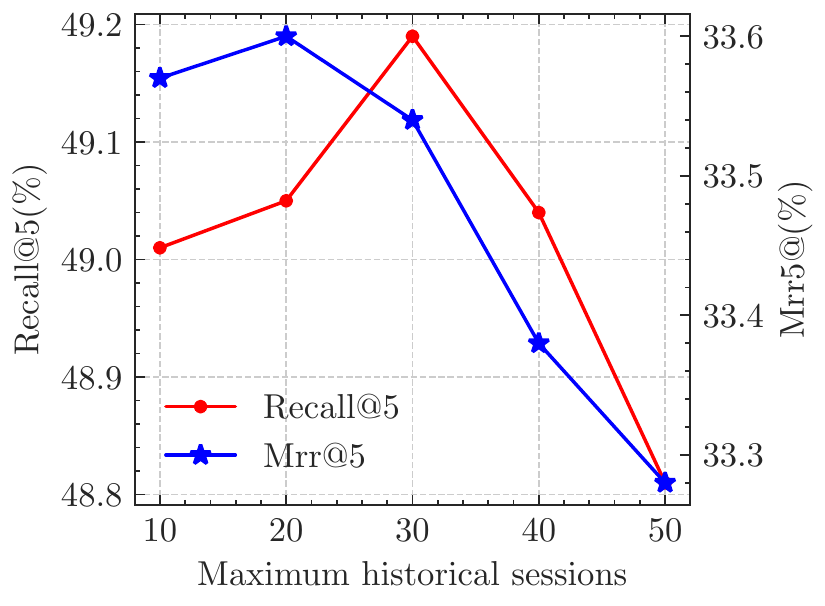}
		\label{fig_second_case}}
	\caption{Performance of A-PGNN in terms of Recall@5 and MRR@5 with different maximum historical sessions.}
	\label{maxsession}
\end{figure}

\begin{figure}[t]
	\centering
	\subfloat[Xing data]{\includegraphics[scale=0.5]{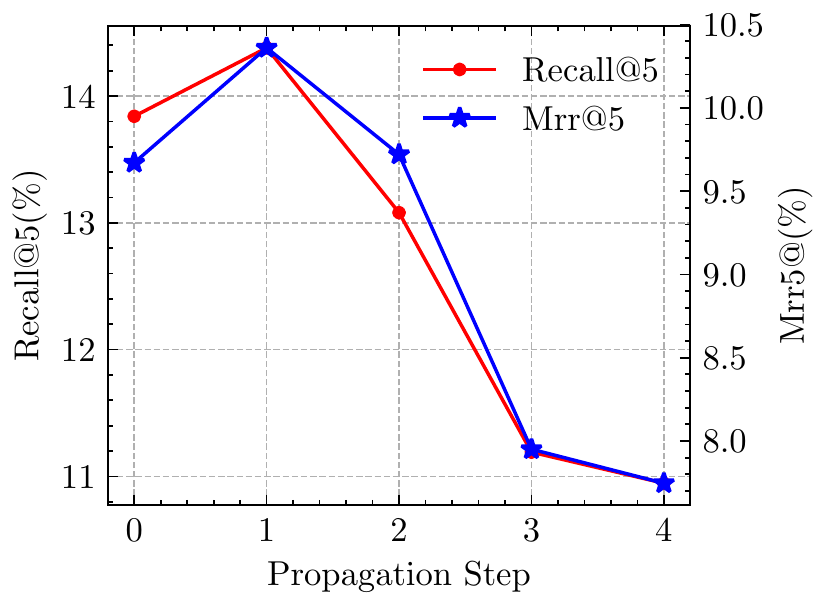}
		\label{fig_first_case}}
	\subfloat[Reddit data]{\includegraphics[scale=0.5]{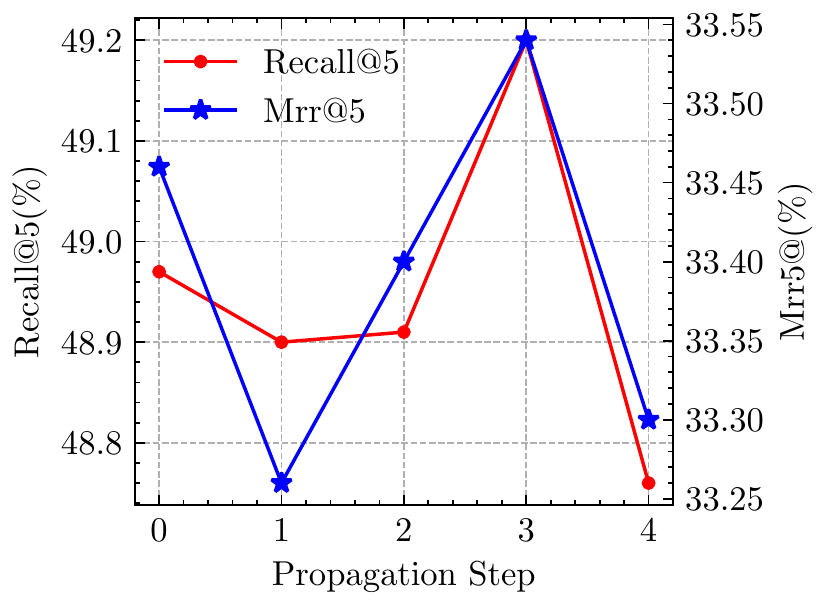}
		\label{fig_second_case}}
	\caption{Performance of A-PGNN in terms of Recall@5 and Mrr@5 with different propagation steps.}
	\label{pgnn_layer}
\end{figure}

\section{Conclusions}
In this paper, we propose PGNN for session-aware recommendation scenario. A-PGNN captures the complex transition relationships between items in each user behavior graph by the PGNN. At the same time, it uses the Dot-Attention mechanism to explicitly model the effect of historical sessions on the current session, which makes it easy to capture the user's long-term performance. Comprehensive experiments on two public data sets verify the effectiveness of different components in our model and confirm that A-PGNN can outperform other state-of-the-art models in most cases. 

For future work, we will improve the flexibility and scalability of PGNN by incorporating the dynamic graph neural networks. Besides, we are also interested in exploring more effective attention mechanisms to integrate the users’ long- and short-term interests.

\ifCLASSOPTIONcaptionsoff
  \newpage
\fi


\bibliographystyle{IEEEtran}
\bibliography{tkde}

\end{document}